\documentclass[runningheads]{llncs}
\usepackage[hidelinks]{hyperref}
\usepackage{subfigure}
\usepackage{graphicx}
\begin{document}
\title{\textit{Making Beshbarmak}: Games for Central Asian Cultural Heritage}
\titlerunning{Cultural Games for Central Asian Heritage}

\author{Amina Kobenova$^{1}$ \and Adina Kaiymova, MBA$^{2}$}
\institute{
    $^{1}$University of California, Santa Cruz, 95064, USA \\
    $^{2}$One League, San Francisco, California, 94108, USA
}

\authorrunning{Kobenova and Kaiymova}

\maketitle              % typeset the header of the contribution
\begin{abstract}
 This paper introduces ``Making Beshbarmak'', an interactive cooking game that celebrates the nomadic ancestry and cultural heritage of Central Asian communities worldwide. Designed to promote cultural appreciation and identity formation, the game invites players to learn and recreate the traditional dish \textit{Beshbarmak} through an engaging step-by-step process, incorporating storytelling elements that explain the cultural significance of the meal. Our project contributes to digital cultural heritage and games research by offering an accessible, open-source prototype on p5.js, enabling users to connect with and explore Central Asian traditions. ``Making Beshbarmak'' serves as both an educational tool and a platform for cultural preservation, fostering a sense of belonging among Central Asian immigrant populations.

 % Our contribution is two-fold: enhancing digital cultural heritage research through gamified interactive media, and providing a working prototype accessible to the global community via p5.js open-source repositories. This project aims to foster cultural appreciation and identity formation among Central Asian populations.

\end{abstract}

\begin{figure}[h]
    \centering
    \subfigure[Cooking Table and Campfire by Yurt in Tian Shan Mountains]{
        \includegraphics[width=0.47\textwidth]{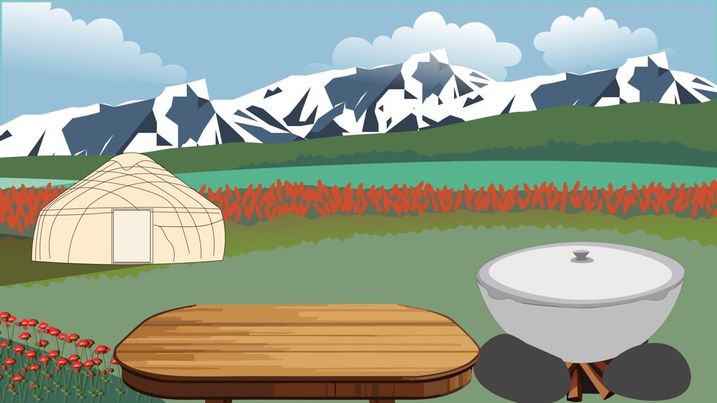}
        \label{fig1}
    }
    \hfill
    \subfigure[Kitchen Scene with Ingredients and Oyu-Decorated Pots]{
        \includegraphics[width=0.47\textwidth]{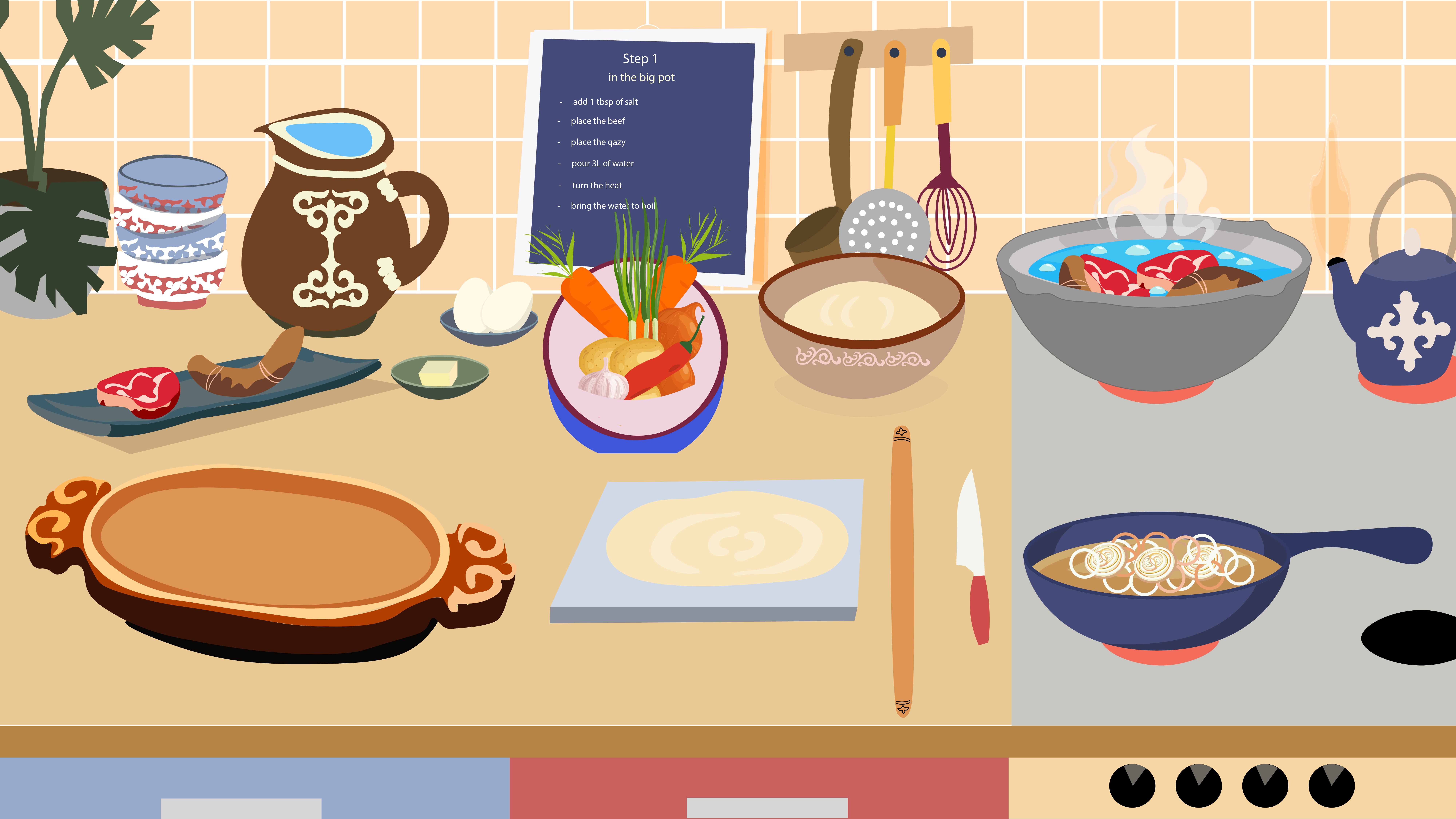}
        \label{fig2}
    }
    \caption{Game Snippets from Making Beshbarmak, Part 1.}
    \label{fig:collage1}
\end{figure}

\section{Introduction}

Thousands of Central Asian immigrants arrive in English-speaking countries annually \cite{noauthor_migration_2024}, seeking a better quality of life, employment and education opportunities. While adult immigrants arrive with significant cultural backgrounds and pre-established identities, children and adolescents are especially vulnerable when risks of building a new life abroad. According to U.S. Health Status and Risk Behaviors of Adolescents \cite{hernandez2008children}, immigrant children experience larger emotional and cultural gaps when acculturating to the new environment, which could lead to higher anxiety and stress levels.

Building a cultural identity is one of the key factors in human development and well-being \cite{ward2023acculturation}, \cite{suh2002culture}. Raising appreciation of one’s historical and cultural backgrounds may yield better results in reducing the acculturation stress and fostering a healthier cultural identity. Game developers, digital media artists, and researchers aim to create experiences that celebrate diversity and promote human flourishing, ranging from acculturative games \cite{conde2023acculturative}, to culture, nature, and postcolonialism-centered interactive installations \cite{dhabia_oasis_2023}, \cite{al-yahya_you_2023}.

To celebrate the nomadic ancestry and cultural background of Central Asian families, we introduce an interactive cooking game in which children and adults can learn to cook the traditional Central Asian meal -- \textit{Beshbarmak}, a festive meat stew popular in Kazakh, Kyrgyz, and Uzbek cuisines \cite{jacob2018international}.

With this project, our contribution to the EAI ArtsIT community is two-fold. First, we contribute to the growing body of research projects on digital cultural heritage, focusing on promoting flourishing and well-being through interactive media and gamification. Second, we propose a working prototype of a game made with the creative coding platform p5.js. The game will be made available to the public via the p5.js platform \cite{p5js} and the open-source community on Github \cite{processingp5js_2024} for any user to play worldwide.

\section{Related Works}

Central Asia occupies around 14.2\% of the world's land \cite{golden2011central} and is the homeland to millions of migrants worldwide \cite{noauthor_migration_2024}. Located between Eastern Europe and East Asia, it consists of 5 countries: Kazakhstan, Uzbekistan, Tajikistan, Kyrgyzstan, and Turkmenistan \cite{golden2011central}. Yet, the cultural heritage in digital media of the region remains largely understudied, and human-computer interaction (HCI) as a discipline is rarely institutionalized. Existing work focuses on post-colonial developments in technology \cite{straeten2022tech} or technological infrastructure \cite{straeten2022tech}. Because of the unique geographical placement, Central Asia rarely falls into exploratory HCI case studies, such as \cite{smith2007institutionalizing} and \cite{sari2015understanding}, or regional HCI conferences, such as \cite{hci4sa}. To the best of our knowledge, our project is one of the first ones to document cultural identity of the region through games and interactive media.

% Acculturative game design is a new term in the digital media community introduced by Conde et al. in late 2023 \cite{conde2023acculturative}. In their work, Conde et al. propose a framework for acculturation games through studying the ``Latine'' immigrant population in North America. They describe ``acculturation stress'' as a phenomenon in which first-generation immigrant children, particularly of Latine origin, face double stress related to barriers of immigration, cultural assimilation, and building personal identity. Conde et al. argue that “serious games provide a unique opportunity to address this challenge” by “introducing novel experiences to encourage the growth of perspectives in acculturative norms” \cite{conde2023acculturative}. Building on Conde et al.’s framework of acculturative games, we propose a creative coding gaming application to provide an environment for Central Asian immigrant youth to learn about their origin through gamifying cultural heritage.

Many games and serious games exist in realm of cultural heritage and education. Anderson et al. \cite{anderson2010developing} provide a systematic review and taxonomy on tangible and gameful experiences to preserve cultural heritage. However, many of the cited games focus on architecture, historical wars, and natural sites, paying little emphasis on the ``soft" aspects of culture, such as food and familial ties. We found three most recent games focusing on sharing food as an element of cultural heritage available on public platforms: \textit{Nainai's Recipe} \cite{tech_nainais_2021}, \textit{Venba} \cite{venba_nodate}, and \textit{Hot Pot For One} \cite{hotpot_nodate}. Both \textit{Nainai's Recipe} \cite{tech_nainais_2021} and \textit{Hot Pot For One} \cite{hotpot_nodate} utilize storytelling and narrative to learn and interact with the Chinese cuisine and culture, while \textit{Venba} focuses on Indian immigrants in Canada and explores themes of ``family, love, and loss'' through its narrative around food \cite{venba_nodate}.

To the best of our knowledge, no games around Central Asian food and heritage exist on public gaming platforms. Inspired by rich opportunities of visual storytelling, interactivity, and play, we propose ``\textit{Making Beshbarmak}'' as a contribution to the growing body of games and media research on Central Asian heritage and a tribute to immigrant families building their lives and identities outside of home.

\section{Game Design and Discussion}

\begin{figure}[h]
    \centering
    \subfigure[Cooked Beshbarmak Dish]{
        \includegraphics[width=0.47\textwidth]{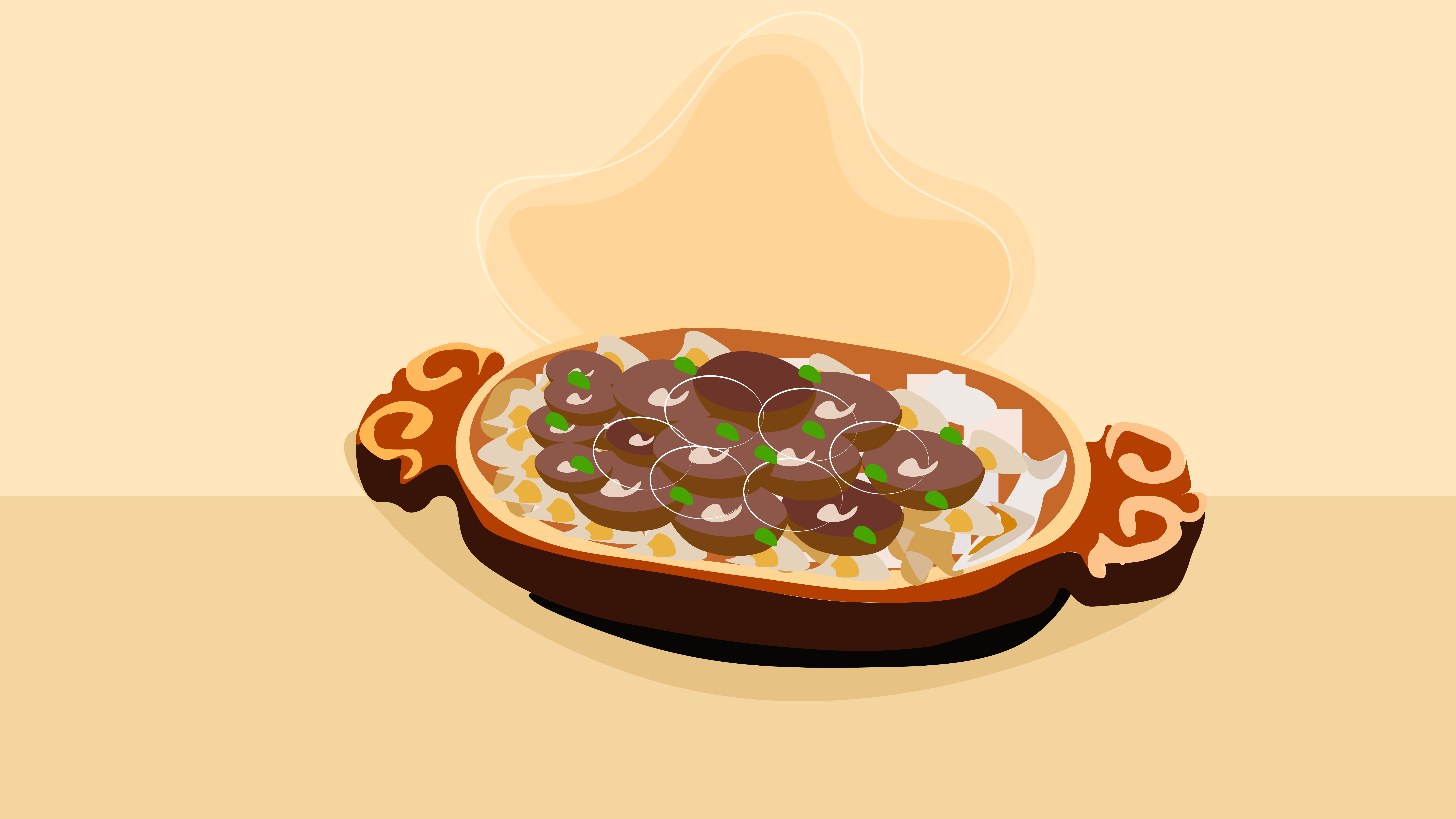}
        \label{fig3}
    }
    \hfill
    \subfigure[Meal and Campfire by Yurt in Tian Shan Mountains]{
        \includegraphics[width=0.47\textwidth]{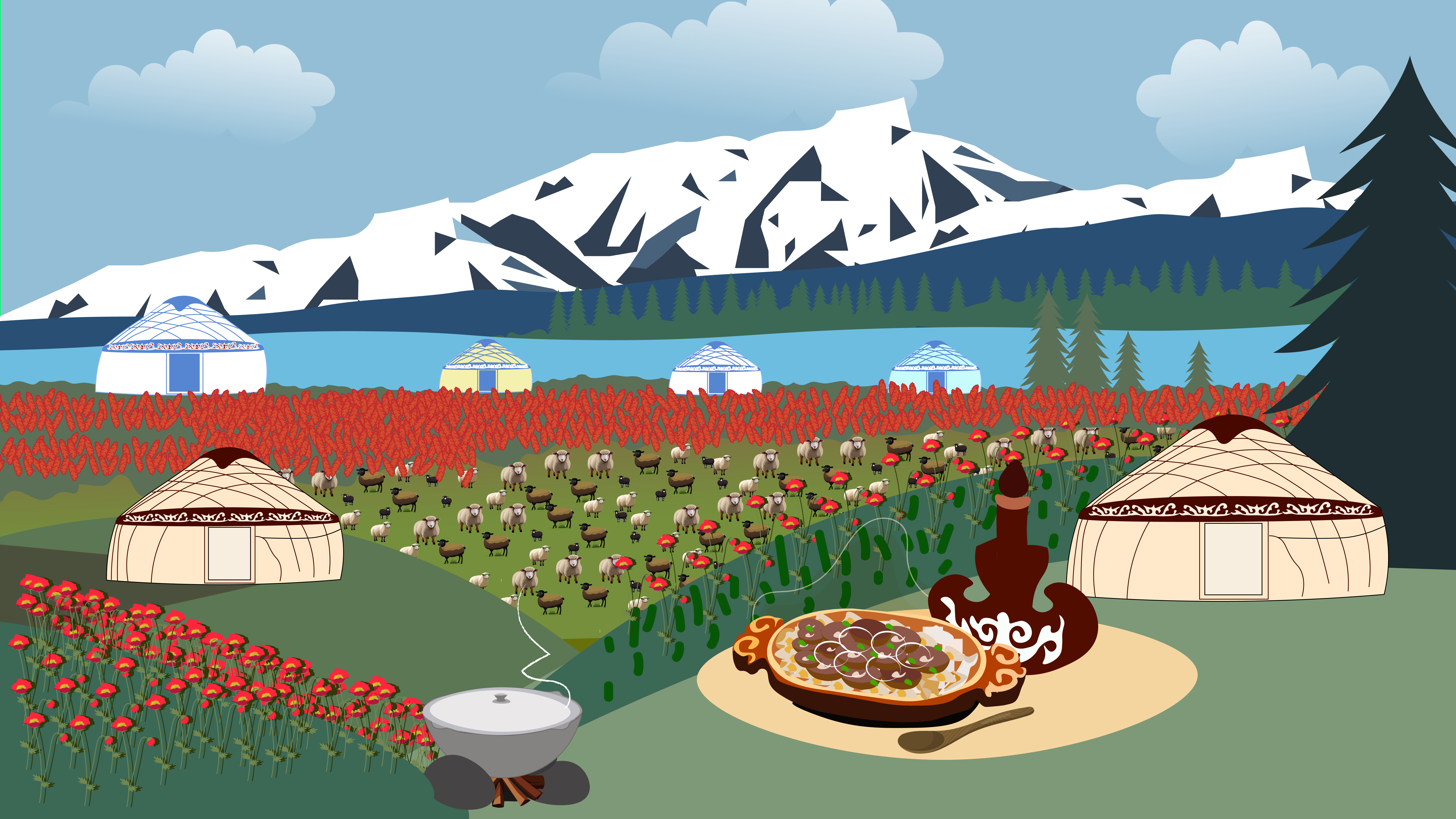}
        \label{fig4}
    }
    \caption{Game Snippets from Making Beshbarmak, Part 2.}
    \label{fig:collage2}
\end{figure}

We propose a web-based 2D game prototype, hosted on a p5.js open-source coding platform. The gameplay emphasizes the cultural learning process by having players follow detailed instructions for each cooking step. As players interact with ingredients through dragging and dropping, they receive contextual prompts about the cultural significance of each ingredient and its role in the dish. For example, the use of particular food products in Kazakh Beshbarmak will be explained, and players can choose between different cultural variations of the dish as seen in Kyrgyz or Uzbek cuisine, deepening their engagement with Central Asian culinary diversity. These interactions reinforce cultural learning by connecting gameplay actions to heritage.

In addition to food, we leverage several aspects of Central Asian culture through our game design: traditional local art \textit{``Oyu''}, natural landscapes, and nomadic houses. Our game starts with a landing scene depicting a traditional nomadic outdoor setting. The player is then prompted to begin the cooking experience and interact with food products to re-create the meal. After creating the meal, the user returns to the outdoor setting with a finished meal. More images depicting game design are available in Appendix Figures \ref{fig1}, \ref{fig2}, \ref{fig3}, \ref{fig4}.

Interactive media and game artifacts provide an ample opportunity for sociology and cultural heritage researchers to explore new ways of engaging with culture through technology. Through continued user studies, these tools can provide new insights and innovative ways of interacting with traditions and cuisine, especially in immigrant populations. ``Making Beshbarmak'' exemplifies how gamification and interactive storytelling can be leveraged to celebrate cultural knowledge and appreciation for diverse cultural backgrounds.

While we create a game project that reveals an aspect of Central Asian heritage, it is important to note that games alone cannot teach what it is to be from a certain culture. Games can be used as a great learning tool to get exposure to certain cultures and as a start to the larger conversation on cultural identities, but they are not a substitute for institutionalized help or psychological interventions for immigrant families and their children. Games are about shared play and telling stories, and this is our hope for this project: to open up a broader conversation on what it means to be of Central Asian descent, anywhere in the world.

Future work should focus on specific user studies in the context of digital cultural heritage and games. This includes developing more sophisticated and immersive interactive media experiences that can further document and express the rich cultural tapestry of Central Asian communities. Additionally, collaborative efforts with cultural experts, psychologists, and educators can enhance the depth and impact of such projects, ensuring they contribute meaningfully to both cultural preservation and the well-being of immigrant populations.

\section{Conclusion}
In this paper, we presented ``Making Beshbarmak,'' an interactive game designed to celebrate and preserve the cultural heritage of Central Asian immigrant families. By focusing on the traditional dish Beshbarmak, the game provides a platform for both children and adults to connect with their cultural roots, promoting a sense of identity and well-being in a new environment. Our work contributes to the field of digital cultural heritage, emphasizing positive cultural identity formation through interactive media artifacts.

\section*{Acknowledgements}
We would like to thank Dr. Sri Kurniawan at UC Santa Cruz, Symbat Bekzhigit at NYU Abu Dhabi, and the Central Asian community in Northern California for their valuable support in the development of this project.


\end{document}